\documentclass[sigconf,screen,authorversion]{acmart} %

\usepackage{subcaption}
\usepackage[skip=3pt]{caption}
\usepackage{booktabs}
\usepackage{graphicx}
\usepackage[normalem]{ulem}
\useunder{\uline}{\ul}{}
\usepackage{amsthm}
\usepackage[inline]{enumitem}
\usepackage[acronym]{glossaries}
\makeglossaries
\usepackage[textsize=tiny]{todonotes}
\usepackage{multirow}
\usepackage[capitalize,noabbrev]{cleveref}
\usepackage{balance}

\theoremstyle{definition}

\AtBeginDocument{%
  \providecommand\BibTeX{{%
    \normalfont B\kern-0.5em{\scshape i\kern-0.25em b}\kern-0.8em\TeX}}}

\newacronym{pgm}{PGM}{Probabilistic Graphical Model}

\newacronym{nn}{NN}{Neural Network}
\newacronym{llm}{LLM}{Large language model}
\newacronym{serp}{SERP}{Search Engine Results Page}
\newacronym{pbm}{PBM}{Position-Based Model}
\newacronym{cm}{CM}{Cascade Model}
\newacronym{ubm}{UBM}{User Browsing Model}
\newacronym{dcm}{DCM}{Dependent Click Model}
\newacronym{dbn}{DBN}{Dynamic Bayesian Network Model}
\newacronym{ncm}{NCM}{Neural Click Model}
\newacronym{csm}{CSM}{Click Sequence Model}
\newacronym{pcsm}{PCSM}{Partially Sequential Click Model}
\newacronym{tcm}{TCM}{Temporal Click Model}
\newacronym{thcm}{THCM}{Temporal Hidden Click Model}
\newacronym{emalgorithm}{EM}{Expectation-Maximization}
\newacronym{ips}{IPS}{Inverse Propensity Scoring}
\newacronym{mle}{MLE}{Maximum Likelihood Estimation}

\makeatletter
\newcommand\headingnodot{\def\@toclevel{4}%
  \@startsection{paragraph}{4}{\z@}%
  {-.2\baselineskip \@plus -2\p@ \@minus -.2\p@}%
  {-3.5\p@}%
  {\ACM@NRadjust{\bfseries}}}
\newcommand{\heading}[1]{\headingnodot{#1.}}
\makeatother

\author{Santiago de Leon-Martinez}
\affiliation{%
  \institution{Brno University of Technology}
  \city{Brno}
  \country{Czechia}
}
\additionalaffiliation{%
  \institution{Kempelen Institute of Intelligent Technologies}
  \city{Bratislava}
  \country{Slovakia}
}
\email{santiago.deleon@kinit.sk}
\orcid{0000-0002-2109-9420}

\author{Jingwei Kang}
\affiliation{%
  \institution{University of Amsterdam}
  \city{Amsterdam}
  \country{The~Netherlands}
}
\email{j.kang@uva.nl}
\orcid{0009-0003-9283-4060}

\author{Robert Moro}
\affiliation{%
  \institution{Kempelen Institute of Intelligent Technologies}
  \city{Bratislava}
  \country{Slovakia}
}
\email{robert.moro@kinit.sk}
\orcid{0000-0002-3052-8290}

\author{Maarten de Rijke}
\orcid{0000-0002-1086-0202}
\affiliation{%
  \institution{University of Amsterdam}
  \city{Amsterdam}
  \country{The~Netherlands}
}
\email{m.derijke@uva.nl}

\author{Branislav Kveton}
\orcid{0000-0002-3965-1367}
\affiliation{%
  \institution{Adobe Research}
  \city{San Jose, CA}
  \country{United States}
}
\email{kveton@adobe.com}

\author{Harrie Oosterhuis}
\orcid{0000-0002-0458-9233}
\affiliation{%
  \institution{Radboud University}
  \city{Nijmegen}
  \country{The~Netherlands}
}
\email{harrie.oosterhuis@ru.nl}

\author{Maria Bielikova}
\orcid{0000-0003-4105-3494}
\affiliation{%
  \institution{Kempelen Institute of Intelligent Technologies}
  \city{Bratislava}
  \country{Slovakia}
}
\email{maria.bielikova@kinit.sk}

\copyrightyear{2025}
\acmYear{2025}
\setcopyright{cc}
\setcctype{by}
\setcctype{by}
\acmConference[SIGIR '25]{Proceedings of the 48th International ACM SIGIR
Conference on Research and Development in Information Retrieval}{July 13--18,
2025}{Padua, Italy}
\acmBooktitle{Proceedings of the 48th International ACM SIGIR Conference on
Research and Development in Information Retrieval (SIGIR '25), July 13--18,
2025, Padua, Italy}
\acmDOI{10.1145/3726302.3730301}
\acmISBN{979-8-4007-1592-1/2025/07}

\ccsdesc[500]{Information systems~Recommender systems}
\ccsdesc[500]{Human-centered computing~User studies}
\ccsdesc[500]{Human-centered computing~Human computer interaction (HCI)}

\keywords{Carousel interfaces, Browsing behavior, Eye tracking}

\begin{document}

\title{RecGaze: The First Eye Tracking and User Interaction Dataset for~Carousel~Interfaces}

\begin{abstract}
Carousel interfaces are widely used in e-com\-merce and streaming services, but little research has been devoted to them.
Previous studies of interfaces for presenting search and recommendation results have focused on single ranked lists, but it appears their results cannot be extrapolated to carousels due to the added complexity.
Eye tracking is a highly informative approach to understanding how users click, yet there are no eye tracking studies concerning carousels.
There are very few interaction datasets on recommenders with carousel interfaces and none that contain gaze data.

We introduce the RecGaze dataset: the first comprehensive feedback dataset on carousels that includes eye tracking results, clicks, cursor movements, and selection explanations. The dataset comprises of interactions from 3 movie selection tasks with $40$ different carousel interfaces per user. In total, 87 users and 3,477 interactions are logged.
In addition to the dataset, its description and possible use cases, we provide results of a survey on carousel design and the first analysis of gaze data on carousels, which reveals a golden triangle or F-pattern browsing behavior.

Our work seeks to advance the field of carousel interfaces by providing the first dataset with eye tracking results on carousels.
In this manner, we provide and encourage an empirical understanding of interactions with carousel interfaces, for building better recommender systems through gaze information, and also encourage the development of gaze-based recommenders. %
\end{abstract}

\maketitle

\section{Introduction}
Single ranked lists have been the default manner for presenting search and recommendation results for many years~\citep{hearst-2009-search}.
Different domains and scenarios may, however, require different interfaces. 
For instance, in web image search results are typically displayed in a 2D grid (i.e., single 2D ranked list), allowing users to inspect the results in a vertical as well as a horizontal direction~\citep{xie-why-2018}.
\emph{Carousel} and \emph{multi-list}\footnote{Carousel interfaces are a subclass of multi-list interfaces with swipeable lists called carousels.} interfaces have become a popular way for e-commerce and streaming services to display items (see \cref{fig:user_study}). 
These interfaces organize items into topics, where a topic consists of items that share certain characteristics. Multiple topics are shown below each other in a list that can be scrolled vertically and each topic is shown as a row that can be scrolled horizontally~\citep{bendada-2020-carousel}.

While carousels are an effective means of displaying recommendations through their use and success in practice (e.g. Netlifx or Spotify), there is little research on why carousels may be better than other presentations formats. Moreover, there is little research in how users interact them, which may greatly impact how systems are designed in practice. 
In contrast to the hundreds of publications on simpler interfaces like single 1D ranked lists and single 2D ranked lists~\citep[see, e.g.,][]{chuklin-2015-click}, we have only been able to locate fewer than 30 publications on carousels and multi-lists. We believe that this lack of research is due to:
\begin{enumerate}[leftmargin=*]
    \item a lack of publicly available datasets of  interaction data with carousels and multi-lists, and 
    \item a lack of empirical studies of users' browsing behaviors.
\end{enumerate}

\noindent%
Our goal in this paper is to address the first problem. We leave addressing the second problem as future work. We hope that the release of a dataset of interaction data with carousels enables researchers to advance the field of carousel and multi-list interfaces.

This work extends beyond just carousel/multi-list interfaces and presents the first publicly available recommender dataset with eye tracking. Eye tracking technology has been steadily developing over the years. Examples of this can be seen in the continued development and popularity of virtual reality (VR) and augmented reality (AR) devices (e.g., Apple Vision Pro~\cite{appleAppleVision} or Meta Quest 3~\cite{meta}), which point to a near future where a VR/AR headset or glasses replace today's devices. 
Gaze will soon be an accessible online data stream for recommender system designers that can revolutionize recommenders by providing item observations, item dwell times, and gaze interaction sequences before a click. The latter can greatly help in determining what leads to a click or purchase. 

This work seeks to achieve the following objectives: 
\begin{enumerate}[leftmargin=*]
    \item Provide the first recommendation dataset with eye tracking data (along with cursor and selection explanations), enabling researchers to empirically understand how users browse carousel interfaces and advance the field of gaze-based recommenders;
    \item Provide the first click feedback dataset in carousels and the second in multi-lists with more than 9 times as many interactions (movie selections) than prior work, enabling research in the underdeveloped field of carousel/multi-list~\cite{10.1145/3450613.3456809}; and
    \item Present the first gaze visualizations in carousel interfaces supporting golden triangle or F-pattern browsing behavior. %
\end{enumerate}

\section{Related Work}

\subsection{User Interactions}
\label{Related Studies on User Interaction}

\heading{Single 1D ranked lists \& search engine results pages}
In traditional single 1D ranked lists and search engine results pages, top-down browsing behavior was originally assumed. It was confirmed empirically through both click data \cite{joachims_accurately_2016} and eye tracking studies, in both search results \cite{granka_eye-tracking_2004, cutrell_what_2007,pan_determinants_2004,hofmann_eye-tracking_2014} and recommender systems \cite{castagnos_eye-tracking_2010, 10.1145/2959100.2959150, li_towards_2017}. These findings of top-down browsing behavior inspired the popular cascade click model \cite{craswell_experimental_2008}. Additionally, eye tracking studies were not only necessary to truly confirm browsing behavior, but also helped to learn parameters of click models. 

\heading{Single 2D ranked lists (slates \& grids)}
For single 2D ranked lists, eye tracking studies were also performed to determine browsing behavior. \citet{10.1145/1743666.1743736} were one of the first to find that users did not browse line by line (left-right or top-down ) and instead used a mixture of both. Later, \citet{10.1145/2959100.2959150} found that in grid-based interfaces, user gaze behavior follows an F-shaped pattern (also known as the ``golden triangle'') from top to bottom and from left to right, rather than being concentrated in the center of the screen.
However, other studies have found that the F-shaped pattern does not apply to image search and browsing based on grid layouts \cite{doi:10.1177/154193120705101831, doi:10.1177/1541931214581234}.
\citet{10.1145/3308558.3313514} further examined user attention on grid-based image search results page and found behavioral patterns such as ``Middle bias,'' ``Slower decay,'' and ``Row skipping.'' As single 2D ranked lists exhibit more complex browsing behaviors than single 1D ranked lists, it is reasonable to expect that carousel interfaces may have even more complex and varied browsing behaviors.

\heading{Multi-list/carousel interfaces}
There are only a few user studies in carousel/multi-list interfaces and no eye tracking studies. \citet{10.1145/3450613.3456809} compared carousels to a single 2D ranked list in item recommendation finding that users were slower and explored longer, while also perceiving the items in carousels as more diverse and novel. Similarly, \citet{starke_serving_2021} compared single 1D ranked lists, single 2D ranked lists, and carousels for recipe recommendations finding that single 2D ranked lists and carousels were both easier to use for participants than the single 1D ranked list. The most recent study by \citet{loepp_how_2023} examined user interactions more completely, finding that item position impacts selection probability (similarly to NDCG2D proposed for carousels \cite{ferrari_dacrema_offline_2022,felicioni_measuring_2021}) and there were noticeable differences in how users perceived and used the carousels. \citet{rahdari_towards_2024, 10.1145/3511095.3531278} simulated users showing the advantages and efficiency of browsing carousels vs.\ single 2D ranked lists to find movies. Additionally, \citet{rahdari_ranked_2022} extended the cascade click model to carousels called the ``carousel click model,'' which assumes that users browse topics until finding one of interest and then examine the items of that topic only.

\subsection{Related Datasets}
There are several commonly used datasets for recommendation systems research, such as the MovieLens dataset~\cite{movielens} or domain-specific ones, such as the MIND news dataset~\citep{wu-2020-mind}. They typically contain large amounts of feedback (clicks and/or ratings), but lack other data sources. For example, we know of only one dataset that contains mouse cursor movements in search \cite{leiva_attentive_2020} and there are no existing datasets with gaze data.

In terms of existing carousel/multi-list recommender datasets, \citet{bendada_carousel_2020} provided a simulation dataset from the music streaming platform Deezer with $n=974,960$ user embeddings and $n=862$ playlist embeddings with ``ground truth'' display-to-stream probabilities. While this is a large dataset, it may not reflect true user behavior and is not a click feedback dataset. To explore the effect of multi-list layout vs. single 2D ranked list in the context of similar-item recommendations, \citet{10.1145/3450613.3456809} conducted a controlled study with 380 interactions (movie selection sessions) on multi-list interfaces with 6 %
unswipeable lists of 5 items each across multiple scenarios varying topics from non-descriptive to ``Movies with the same genre'' or actor/director among others. This is the only click feedback dataset available for multi-list interfaces and it does not include topic labels. Our dataset is more descriptive, is the only available in carousels (swipeable lists), and provides more than 9 times as many interactions.

\section{User Study Methodology}
To better understand how users browse carousel interfaces, we designed an eye tracked user study of 40 screens, where participants pick one movie  found on a carousel page (or exit without selection). The user study was designed with the following goals in mind:
\begin{enumerate}[leftmargin=*]
    \item Determine any general browsing behaviors across users;
    \item Determine the impact of carousel topic (genre) preference on browsing;
    \item Examine the differences in browsing behavior across 3 browsing tasks: free-browsing, semi-free-browsing, direct search;
    \item Survey users to address open questions in carousel design \cite{loepp_multi-list_2023};
    \item Provide an extensive mouse, click, and gaze dataset along with user explanations for future research.
\end{enumerate}

\noindent%
The study consisted of three parts: 
\begin{enumerate*}[label=(\roman*)]
    \item a pre-survey, 
    \item movie selection tasks, and 
    \item a post-survey. 
\end{enumerate*}
The survey sections gather user information, particularly their genre preferences, ideas on carousel interfaces, and feedback on the interface. The main part of the study are 40 screens of movie selection tasks. Of these 40 screens, 30 are free-browsing (pick any movie on the screen), 5 are semi-free browsing (pick a movie from your favorite genre), and 5 are direct search tasks (find a specific movie). The same screens and tasks were used for every participant. We used a larger number of free-browsing screens as this scenario is the closest to real-world interactions and while examining behavior in other scenarios may be insightful, we did not want to make the study too long.

Additionally, after the free-browsing and semi-free browsing search tasks, participants were asked to give feedback on their movie selection: familiarity with the movie selected and why they selected it. Along with selection explanations, we gathered mouse movements, clicks, gaze, and screen recordings throughout the movie selection tasks to provide the most complete recommender system dataset that is publicly available. A summary of the tasks and screens with the number of interactions (screens where feedback data was gathered) can be found in \cref{tab:Click}.

\begin{table}[]
\caption{Task summaries of interaction feedback, the number of movies selected (users can exit without selecting) and average screen time (until selection or exit).}
\label{tab:Click}
\centering
\setlength{\tabcolsep}{0.7mm}
\begin{tabular}{@{}l@{}cc cc c@{}}
\toprule
& &  && Movie& \multicolumn{1}{c}{Avg. screen}
\\
\textbf{Task} &
  \multicolumn{1}{l}{Users} &
  \multicolumn{1}{l}{Screens} & Interactions&
  \multicolumn{1}{c}{Selections}  &
  \multicolumn{1}{c}{time} \\ 
\midrule
Free-browsing        & 87 & 30  & 2,607 & 2,432 & 56.99s \\
Semi-free & 87 & \phantom{0}5   & \phantom{0,}435& \phantom{0,}383  & 43.85s \\
Direct Search & 87& \phantom{0}5   & \phantom{0,}435& \phantom{0,}424             & 11.57s \\ \midrule
All Tasks     & 87 & 40  &3,477& 3,239 & 49.67s \\ \bottomrule
\end{tabular}%
\end{table}

\subsection{Screen Design \& Movie Selection}
Due to eye tracking limitations (physical setup, calibration, accuracy problems with movement, etc.) and the challenges of gathering other forms of feedback (clicks, cursor movements, etc.), we designed a user study in a controlled environment. We created custom screens emulating a carousel experience allowing for feedback to be gathered using available software and the Chrome browser.

\begin{figure*}[t]
    \centering
    \includegraphics[width=0.8\textwidth]{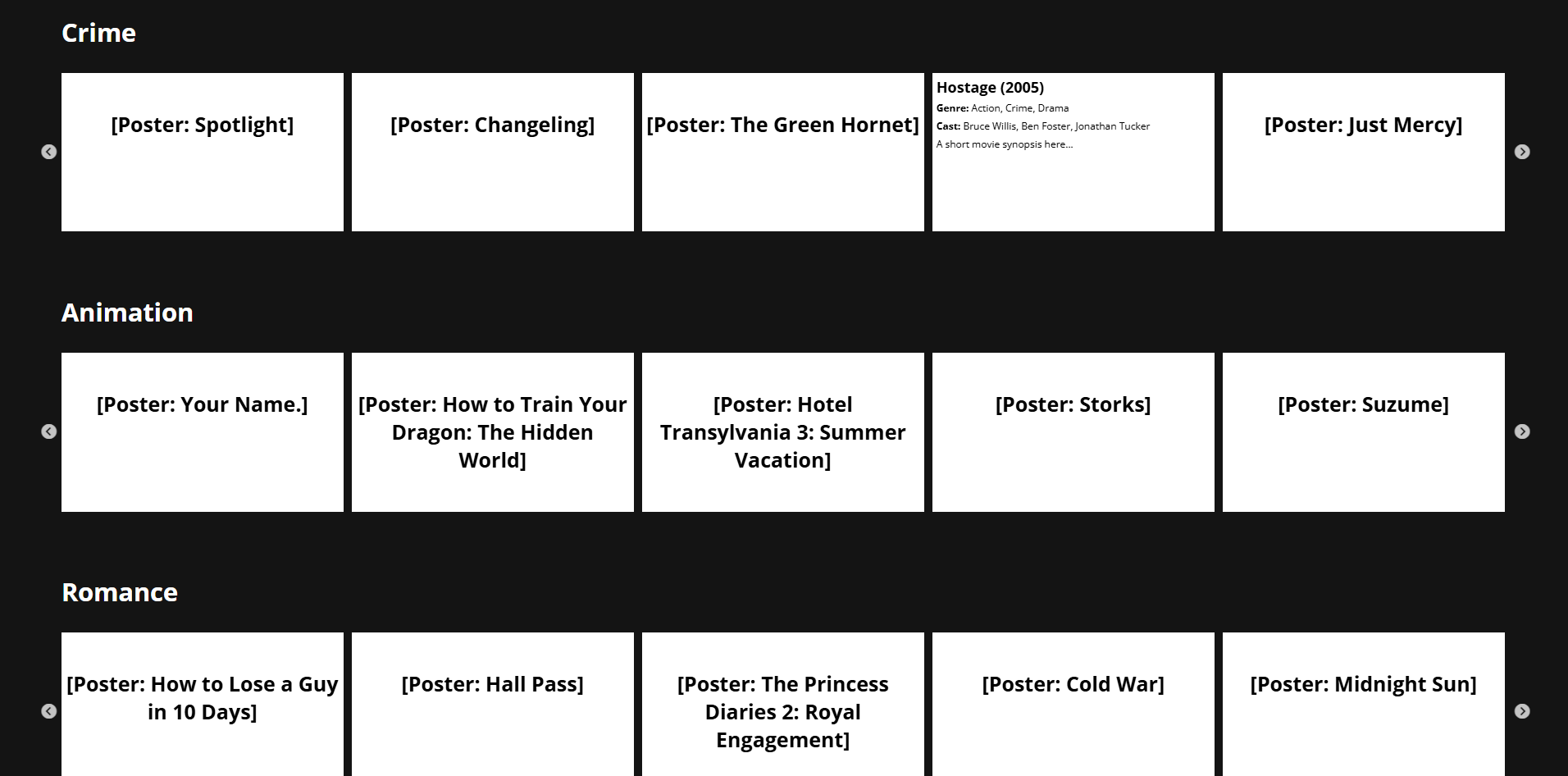}
    \caption{A capture of a free-browsing  screen at initial presentation with the first 3 carousels shown and mouse-over details of a movie in the top row. Movies are ranked (left to right) by popularity based on IMDB number of votes. Movie posters and mouse-over details were shown (removed here due to copyright). See GitHub below for a sample screen recording gif.} %
    \label{fig:user_study}
\end{figure*}

The screens were made to allow participants to browse and select a movie in a carousel interface. Our design was inspired by the Netflix homepage as it is one of the most well-known examples of a carousel interface. This made it easy for participants to be comfortable with the interface and rely on previously learned browsing habits/behaviors. %
In \cref{fig:user_study},  we show a sample screen.

We restrict ourselves to movies from the years 2000 to 2023 to avoid any biases against older movies. Also, we did not include any movie duplicates across/within the screens. %
To provide a suitable number of movies for users to browse and allow at least 2 swipes to be possible (for examining list depth behaviors), we set our carousels to have 15 movies total with 5 movies shown at a time and 3 genre carousels at initial presentation (seen without scrolling down). 

Only genres were used as carousel topics, as one of the goals of the study was to determine how carousel topic preference impacts browsing behavior from the gathered preference information. Genres are a simple and familiar topic for users that are much easier to compare between each other. %

For the genres themselves, we chose 10 genres available from the International Movie Database (IMDB) genres: Action, Animation, Comedy, Crime, Drama, Fantasy, Horror, Romance, Sci-Fi, and Thriller. These 10 cover a wide range of genres that are familiar to users and can be typically seen in Netflix or other streaming services. Moreover, the list contains genres that are likely to be attractive to different user populations, such as Animation and Horror. Strong genre preferences (positive or negative) may have a greater impact on how users browse carousel interfaces. %

\heading{Populating carousels with movies} 
In total, 40 screens were created with 150 movies per screen (15 movies for each of 10 genres). The IMDB non-commercial dataset (extracted on October 3, 2024)\footnote{\url{https://developer.imdb.com/non-commercial-datasets/}} was used for finding top/popular movies to populate our carousels along with The Movie Database (TMDB) API\footnote{\url{https://www.themoviedb.org}} for finding movie posters and movie information (short description, year, actors). We used the number of votes from the IMDB dataset to determine the ranking of the movies (from most votes to least). The top 150 movies were removed to avoid very popular movies that users have seen or would know about (and to provide an instruction screen). We detail the whole procedure of populating carousels with movies below:
\begin{enumerate}[leftmargin=*]
    \item Sort IMDB dataset by number of votes.
    \item Keep only movies from 2000 to 2023.
    \item Remove top 150 movies.
    \item Populate 10 ranked genre movie lists:
    \begin{enumerate}
        \item Take the top movie in the dataset and place it in the top, unfilled position (i.e. append) in a genre list to which it belongs.
        \item Movies with multiple genres are appended to the least filled genre (i.e. highest ranked position among genres), and if genre lists are equally full, then it is decided by random assignment. %
        \item Movies annotated as not having English text in their poster image are excluded.\footnote{Despite this, a few posters did not have text (see limitations in \cref{Limitations})}
    \end{enumerate}
    \item Populate ranked genre carousels for the 40 screens:
    \begin{enumerate}
        \item For each genre, take the top 40 movies of the genre list.
        \item Randomly append these movies to the same top, unfilled position in the 40 genre carousels across all screens (i.e., the 1st movie position for all 40 Horror carousels).
        \item Continue with the next 40 top movies and unfilled positions until filling all genre carousels across the 40 screens.
    \end{enumerate}    
\end{enumerate}

\noindent%
This procedure creates screens with carousels that present 15 movies going from most popular to least popular. Additionally, across the 40 screens each genre carousel is popularity balanced by the random allocation allowing for a fair comparison across screens. 

\heading{Genre ordering} The order of the genre carousels was performed randomly and separately for each of the 3 browsing tasks. For the 30 screens of free-browsing, genre ordering was determined randomly, while also guaranteeing that:
\begin{enumerate*}[label=(\roman*),leftmargin=*]
    \item across the 30 screens, each genre is equally present within three initial presentation locations; and 
    \item across the 30 screens, each genre is equally present in each of the 10 possible positions on the screens. 
\end{enumerate*}
For the 5 screens of semi-free-browsing and 5 screens of  direct search, only (i) above is guaranteed across the 5 screens due to the low number of screens.

\heading{Movie details} 
Movie details are displayed by hovering over a poster with the mouse cursor showing: English title, year, genres, top 3 actors, and a short description. The title, genres, and year are from the IMDB Dataset, while the top 3 actors and the short description are from the TMDB API.

\subsection{Data Collection \& Study Procedure}
\heading{Setting and apparatus} 
Data was collected in Bratislava, Slovakia and Amsterdam, Netherlands using identical setups: a Lenovo Thinkpad T490 laptop, Tobii 4C remote bar eye tracker (90~Hz sampling rate and error of less than 1 degree~\cite{taore_limits_2024,gibaldi_evaluation_2017}), and a 27 inch external monitor with resolution 1920x1080 and window scale 100\%. Gaze, fixation, click, scroll, and cursor data were gathered by Eye Square's in-house testplayer software for eye tracking experiments. In addition, the positions of the browser elements, clicks, and scrolls were also collected through the Chrome browser. Eye tracking data, along with cursor and click data, were analyzed and visualized in Blickshift Analytics Build 2024, sub version 6 \cite{Blickshift_2024}. 

\heading{Study sample} 
Participants were gathered either from the institutions themselves or a coworking space, with those from the coworking being offered remuneration through a 15 euro coupon. The eligibility criteria were: 
\begin{enumerate*}[label=(\roman*)]
\item 18 years of age or older and 
\item ability to use mouse and keyboard to navigate. 
\end{enumerate*}
All participants were first given an explanation of the study along with an informed consent for participation in the study and for data sharing and processing. 

In total, 87 participants were included in the final dataset, of which 47 were men and 39 woman with the following ages: (18--19): 1, (20--29): 50, (30--39): 29, (40--49): 6, and (60--69): 1. 61 participants were gathered in Bratislava while the remaining 26 were in Amsterdam. Additionally, 68 of the participants reported that they had used Netflix or a similar streaming service before; 19 had not.

\heading{Study procedure} All participants consented and fully completed the study in one-sitting composed of 3 parts as described below. 

\heading{Pre-survey} After informed consent, participants were given a pre-survey that gathered basic information (personal information, age, gender) of the participants. Genre preference was assessed through 3 different questions: marking preferred/non-preferred for each, 1-star to 5-star rating for each, %
and pick the top or most preferred genre. The 3 different genre preference questions cover a wide range of common feedback, for comparison among each other (ranking genre preference) and to be relatable to other feedback datasets. To avoid any positional bias in survey presentation, the order of the genres were randomized for the preference questions. 

Along with genre preference, information on frequency of use of movie/TV streaming services and frequency of movie watching (in any format) were gathered as proxies of user expertise with carousel interfaces and movie expertise.  

As a follow-up to a previous user study into carousel interfaces by \citet{loepp_how_2023}, we also include a psychological decision-making scale to relate the scores to browsing behaviors in a future work. We also seek to help address the open questions presented in the same user study and survey paper \cite{loepp_multi-list_2023}: what are the best type of topics to use for carousels? For 10 different types of carousels (i.e., genre-based, content-based, item-based, etc.), we ask users which they would find most helpful (see \cref{survey most helpful topics}). %

\heading{Movie selection tasks} The movie selection tasks were administered in sets of 10 screens each beginning by calibrating the eye tracker. Re-calibration of the eye tracker can improve the accuracy of the tracking, especially for tasks that can take many minutes to complete. The free-browsing task was separated into 3 sets of 10 screens (for a total of 30 screens) and the 5 screens of semi-free browsing were followed by 5 screens of direct search to complete the last set of 10 screens. Half of the participants completed the 3 sets of 10 free-browsing screens followed by the 5+5 screens of the semi-free-browsing and direct search, while the other half completed the semi-free-browsing and direct search set first. This was done to control for the effect of tiredness from previous screens. %

Across all tasks, participants were asked to imagine that they were at home and had opened Neftlix (or a similar streaming service) to find a movie to watch. %
In the case of the free-browsing task, they were asked to pick any movie that they would be interested in watching. In the semi-free-browsing task, participants were simply limited to picking a movie from their top genre choice from the pre-survey. In the direct search task, participants were instructed to find a specific movie and also provided the genre in which the movie was found (see GitHub for a visualization of search targets). The 3 tasks cover different scenarios/intents a user may have when finding a movie using a carousel homepage (without use of the search); we will examine the differences in the eye tracked browsing behavior in our future work. 

With regards to study timing, in general we did not limit the time participants spent on the screens. For the free-browsing task, on average participants spent 56.99 seconds on a screen with a minimum of 1.17s and maximum of 412.09s. However, some participants who spent 2 minutes or more on the majority of free-browsing screens were advised to try to browse a bit faster and that they only needed to pick a movie they would like to watch. These participants that were advised (and when) are marked in the dataset.

\heading{Post-movie selection explanation} 
Following movie selection in free-browsing and semi-free-brows\-ing screens, two questions were asked to determine the users' familiarity with the movie (i.e., already watched, seen a trailer, heard of the movie, unheard of, etc.)  and why they picked the movie (i.e., poster, details, already wanted to watch, movie has been  recommended, I don't know, etc.). %

\heading{Post-survey} After completion of the 40 screens of movie selection, participants filled out a post-survey. The post-survey gathered information on the user experience (from 1--10) of the free-browsing task and tiredness/exhaustion after completing the study. %
Additionally, we gathered feedback on the interface design asking if the participants were overwhelmed by the number of genres or number of movies in each carousel and if they would like to see more movies or genres along with different types of topics for carousels (see \cref{Surveys on carousel design}). The last two questions provided  feedback/suggestions for improving the interface and if participants would be willing to customize their carousel homepages (for both see \cref{Surveys on carousel design}).

\section{Dataset Description}

In this section we describe the main features, format, and preprocessing of the RecGaze dataset.

\subsection{Dataset Features}
\label{Dataset Features}

We provide two datasets for the community to use due to the sensitive nature of raw gaze and cursor data that could be used to potentially link participants between datasets. The first is a public dataset that can be found on Zenodo\footnote{
\url{https://zenodo.org/records/15270518}} 
with summarized fixation and mouse information with already timestamped labeling of Areas of Interest (AOIs) for cursor, clicks, and fixations. The supplementary material (stimuli, all screen layouts, etc.) are available in a GitHub repository.\footnote{\url{https://github.com/santideleon/RecGaze_Dataset}} %
The public dataset is designed to be an already preprocessed dataset that can be used out of the box for research in recommender systems, information retrieval and human computer interaction. It is organized in 4 dataframes in csv format. The user features, item features, and summary feedback dataframes contain the features (as columns) mentioned in \cref{tab:dataset_features}. We additionally provide a fourth dataframe that is a simple click feedback dataset with only the final movie selection click  %
per screen per user (in total 3,239 selections) and also contains the selection explanation information. 

\begin{table*}[h]
\centering
\caption{Major features of the RecGaze Dataset. Features marked with * are part of the non-public dataset and square brackets {[}{]} refer to multiple responses. To be concise, we use shorthand names, combine similar features, and denote different columns using bold. For a detailed explanation of the features refer to the GitHub documentation.}
\label{tab:dataset_features}
\resizebox{0.95\textwidth}{!}{%
\begin{tabular}{llll}
\toprule
\bf Category &
  \bf Feature name &
  \bf Type &
  \bf Example \\ 
\midrule
\begin{tabular}[c]{@{}l@{}}User \\ features\end{tabular} &
  \begin{tabular}[c]{@{}l@{}}UserID\\ Age* \\ Gender* \\ Genre preferences\\ Interface expertise\\ Movie expertise\\ Decision-making\\ Genre/Movie \#\end{tabular} &
  \begin{tabular}[c]{@{}l@{}}Multi\\ Range\\ Text\\ Multi\\ Multi\\ Multi\\ Int 1-7\\ Text\end{tabular} &
  \begin{tabular}[c]{@{}l@{}}KInIT\_18 (Location\_ID)\\ 30-39\\ Man\\ \textbf{Q\_Top:} Action; \textbf{Q\_Preferred:} {[}Action, Comedy{]}; \textbf{Q\_Ratings 1-5:} {[}Action\_5, Animation\_1...{]}\\ 3-4 times per week\\ 1-2 times per month\\ \textbf{Q\_1:} 1; \textbf{Q\_2:} 3; \textbf{Q\_3:} 5; \textbf{Q\_4:} 7; \textbf{Q\_5:} 4; \textbf{Q\_6:} 5;\\ \textbf{Q\_1:} Yes, I felt overwhelmed by \# of Genres; \textbf{Q\_2:} No, I didn't feel overwhelmed by \# of movies\end{tabular} \\ 
\midrule
\begin{tabular}[c]{@{}l@{}}Item \\ features\end{tabular} &
  \begin{tabular}[c]{@{}l@{}}MovieID\\ Screen position\\ Title \& Year\\ Description \\ Poster\end{tabular} &
  \begin{tabular}[c]{@{}l@{}}Int\\ Multi\\ Multi\\ Text\\ Img\end{tabular} &
  \begin{tabular}[c]{@{}l@{}}372058 (ID from TMDB)\\ \textbf{TaskID/Screen}: 7; \textbf{Carousel\_position}: 1; \textbf{Genre}: Horror; \textbf{Movie\_position}: 6 \\ \textbf{Title}: Hostage; \textbf{Year}: 2005\\ "When a mafia accountant is taken host..."\\ http://\{url\}\end{tabular} \\
\midrule
\begin{tabular}[c]{@{}l@{}}Feedback \&\\ AOI label\\ \end{tabular} &
  \begin{tabular}[c]{@{}l@{}}Gaze (90 Hz)*\\ Fixation (90 Hz)*\\ Cursor (event)*\\ Click (event)*\end{tabular} &
  \begin{tabular}[c]{@{}l@{}}Multi\\ Multi\\ Multi\\ Multi\end{tabular} &
  \begin{tabular}[c]{@{}l@{}}\textbf{Time:} 1.241s; \textbf{Position}: 120, 300 px;\\ %
  \textbf{Time:} 1.241s; \textbf{Position}: 120, 300 px; \textbf{AOI:} Movie\_372058; \textbf{Duration}: 0.891s\\ \textbf{Time:} 2.450s; \textbf{Position}: 100, 600 px; \textbf{AOI:} GenreText\_1\_Horror\\ \textbf{Time:} 10.903s; \textbf{Position}: 100, 142 px; \textbf{AOI:} BackSwipe\_Genre\_1\_Horror\end{tabular} \\ 
\midrule
\begin{tabular}[c]{@{}l@{}}Summary\\ feedback\end{tabular} &
  \begin{tabular}[c]{@{}l@{}} Fixation sequence\\ Cursor sequence\\ Click sequence\end{tabular} &
  \begin{tabular}[c]{@{}l@{}} Multi\\ Multi\\ Multi\end{tabular} &
  \begin{tabular}[c]{@{}l@{}} \textbf{Time:} 1.241s; \textbf{AOI:} Movie\_372058; \textbf{Duration}: 0.891s\\ \textbf{Time:} 2.450s; \textbf{AOI:} GenreText\_1\_Horror; \textbf{Duration}: 1.902s\\ \textbf{Time:} 10.903s; \textbf{AOI:} BackSwipe\_Genre\_1\_Horror;\end{tabular} \\ 
\midrule
\begin{tabular}[c]{@{}l@{}}Selection \\ explanation\end{tabular} &
  \begin{tabular}[c]{@{}l@{}}Movie familiarity\\ Why selected?\end{tabular} &
  \begin{tabular}[c]{@{}l@{}}Text\\ Text\end{tabular} &
  \begin{tabular}[c]{@{}l@{}}I have seen part of the movie\\ {[}Because of the poster, Because of the details, Other: "I like the director"{]}\end{tabular} \\
\midrule
Other  &
  Screen recording* &
  Video &
  25 fps .avi video recording \\ 
\bottomrule
\end{tabular}%
}
\end{table*}

The second is a non-public dataset that contains sensitive information that can be provided by reaching out to the authors. It is better suited for researchers that are more interested in  eye/mouse tracking research or would like to do the gaze preprocessing (e.g., AOI linking and fixations) themselves. We provide example use cases of the dataset in \cref{Use Cases}. In the following section, we outline our preprocessing procedure for AOI determination. %

\subsection{Data Preprocessing}
\heading{Time alignment} All timestamped data (gaze, fixation, cursor, click) is aligned to the screen recording video start time of 0:00. Data before the carousel interface webpage loads as well as after the movie selection click was removed. 

\heading{Gaze preprocessing} All gaze that was found to be outside of the screen was removed from the dataset and thus did not impact fixation calculation. Fixations were calculated using the Velocity-Threshold (I-VT) algorithm \cite{10.1145/355017.355028}. All fixations under 60ms were discarded, a common practice in processing gaze data that removes eye movement artifacts not linked to cognitive processes \cite{foucher_unveiling_2024, rayner_eye_1998}.

\heading{Linking gaze and cursor location with AOIs} One of the most important steps in processing the data to leverage the cursor and eye tracking information, was linking gaze and cursor $(x,y)$ screen pixel location with the corresponding AOI: carousel genre text, movie posters, and swipe buttons. In most eye tracking experiments, AOIs are predetermined using a fixed stimuli (the scene or screen being observed) or computer vision is used for object detection. In our case, we had a moving stimuli (the webpage with carousels), which could be scrolled vertically or swiped (scrolled horizontally).

To adjust for vertical scrolling, a fixed stimuli of the whole webpage  was created using the position of the AOIs scraped as elements from the browser (see \cref{fig:vis_fixation}). Then scroll data was used to adjust $(x,y)$ positions for fixations, cursor movement, and clicks.

To adjust for swiping of the carousels (horizontal scroll), %
click events on the swipe buttons were gathered and then used to signal the change of AOIs to the next or previous set of movies. On the stimuli, 2 more identical pages were added for the 2 possible swipes. Combining the stimuli and the adjustments for vertical scroll and swipes, provides perfect AOI linking for cursor and click data and a possible approach for fixation AOI linking.

\heading{Gaze margin of error and fixation AOIs}
Unlike cursor data, gaze data always has a margin of error in the $(x,y)$ coordinate position.
This is mitigated by the fixation algorithm that has a margin for aggregating gaze within a certain time and position window. Despite this, we provide 2 different methods for fixation AOI processing, one of which assumes that participants are always gazing at an AOI on the screen (if within a distance threshold). 

The first is the most simple and uses the exact rectangular bounds of the AOI in the webpage. In this case, fixation data that fall out of the rectangular bounds of the AOI (e.g., a movie poster) is not attributed to any AOI. %

The second method works upon the first and allocates all unassigned fixation data to the nearest AOI (movie poster, forward or backward swipe button, or carousel genre text) up to a distance of 60 pixels.\footnote{Calculated using the 1 degree error and screen dimensions along with the maximum recorded participant distance to screen 76cm.} This method assumes that a participant is more likely observing an AOI rather than the empty background. Initial gaze analyses showed little difference between the two methods. %

\heading{Data quality and missing values}
The data gathered by the user study for the RecGaze dataset combines many modalities to make the most complete recommender dataset to date. %
To ensure the data quality gathered, we manually examined all fixation scan paths for every participant and screen (e.g., \cref{fig:fixation_subim_fixation_scan}) to check for errors. %

Across all screens there were only 3 free-browsing screens (0.1\%) that failed to collect any feedback (clicks, cursor, gaze, fixation are all missing completely). For cursor data, 55 screens (1.6\%) did not gather any mouse movements. For gaze data, 12 screens (0.33\%) did not collect gaze data. Moreover, for the non-missing gaze data, we estimated the percentage of missing gaze data (below 90 Hz) was 2.9\%. For fixation data, 61 (1.8\%) screens had no fixations calculated.

For click data, we used the browser collected clicks to fill missing data from the software. As it is possible to not click, we compared the click data to the selection explanations to determine screens that were missing the final movie selection click. Only 4 (0.1\%) screens were missing the movie selection. Finally, we reviewed the direct search tasks and found that in 7 screens (1.6\% of direct search screens), the participant failed to successfully find and click the instructed movie. These screens are not included in the click summary dataset and are noted in the GitHub documentation.  

\subsection{Use Cases for RecGaze Dataset}
\label{Use Cases}
The RecGaze dataset is the first recommender dataset to provide a wide range of feedback data (eye tracking, cursor, clicks, and selection explanations). For this reason, it can lend itself to many applications. First and foremost, it allows one to comprehensively analyze and understand possible user behaviors within the carousel interfaces by combining all available modalities as illustrated in~\cref{results}. Secondly, it can be used to fine-tune designs of these interfaces and their underlying models. The following are additional possible use cases for which the dataset may serve:\footnote{When applying the dataset in specific settings, it is worth considering the possible limitations of the dataset as discussed in detail in~\cref{Limitations}.}

\heading{Click models for carousel interfaces} Designing click models based on observed browsing behaviors like the F-pattern in \cref{fig:fixation_subim_aggregate} and using the data to determine model parameters.
\heading{Offline evaluation} The browsing behaviors observed in the fixation or gaze data can be used to design better simulations of carousel users for offline experiments. Also, it can be used to understand positional bias in carousels. %
\heading{Whole page optimization} As an extension of click models or with general insights found in the data, one can better understand how different carousels interact with each other, particularly with the included preference information, and design better page layouts. 
\heading{Impressions} The dataset can also be used for impressions, as they can be extracted and compared to the ``ground truth'' gaze data.

\section{User Study Results}
\label{results}
In this section we describe part of the results of the user study, namely the selection explanations and surveys motivated by the open questions posed by the survey paper of \citet{loepp_multi-list_2023} on how to design carousel interfaces. We also provide the first ever visualizations of gaze data to demonstrate the dataset's utility for analyzing the users' browsing behavior. A more extensive analysis of this behavior is a subject of our future work.  %

\subsection{Familiarity and Explanation for Selection }
\label{movie familiarity}

\begin{figure}[t]
    \centering
    \includegraphics[clip, trim=5mm 10mm 0mm 10mm, width=\columnwidth]{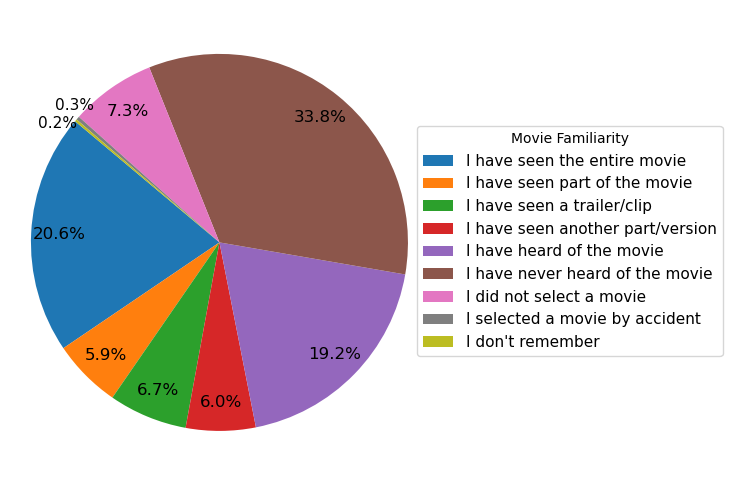}
    \caption{How familiar users are with the selected movie.}
    \label{fig:Movie_Familiarity}
\end{figure}

\heading{User familiarity with movie} 
\cref{fig:Movie_Familiarity} shows the results of the movie familiarity question shown after each of the 3,045 free-browsing and semi-free browsing screens. Only one response was allowed and the question instructed participants to pick the first (topmost) true statement (ordering seen in legend).\footnote{It is important to note that users were instructed to pick any movie that they would like to watch, which may include movies already watched. }

\begin{figure}[t]
    \centering
    \includegraphics[clip, trim=1mm 0mm 1mm 1mm,width=\columnwidth]{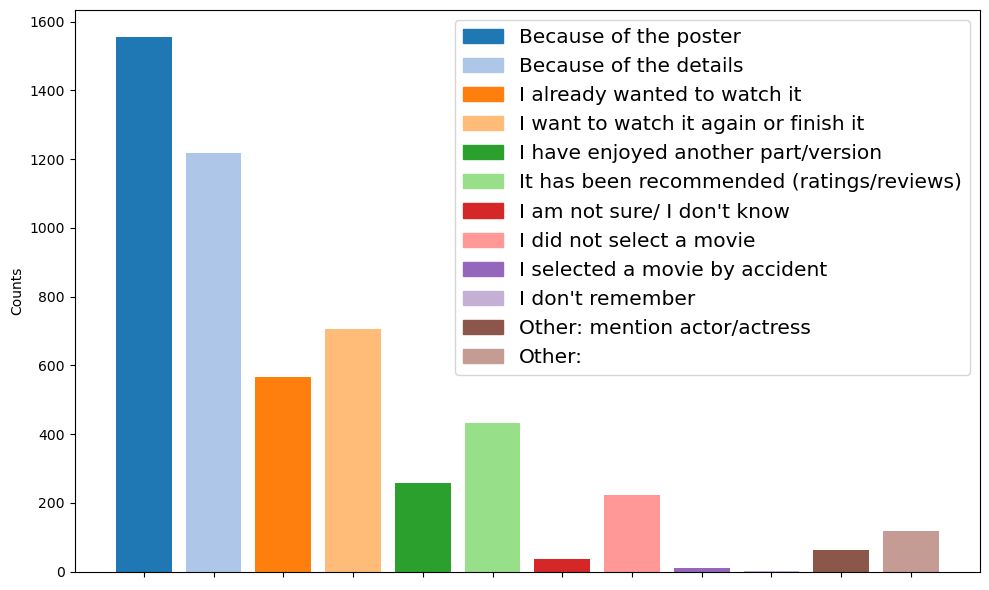}
    \caption{Why users select a movie (multiple responses).}
    \label{fig:Why_picked}
\end{figure}

\heading{Why users selected a movie}
In \cref{fig:Why_picked}, we present the results of the follow-up selection explanation question. %
Participants were able to select multiple responses and allowed to add their own free-form response (other). It can be seen that in more than half of movie selections, participants attributed it to the poster. This points to the importance of item visual biases in carousel interfaces and particularly for the case of movie recommendation.

\subsection{Surveys on Carousel Design}
\label{Surveys on carousel design}
In the only survey paper on carousels and multi-list interfaces, \citet{loepp_multi-list_2023} present the following open questions that we address in this paper: %
\begin{enumerate*}[label=(\roman*)]
\item which types of carousels are preferred by which users, 
\item how many carousels they want to explore, and 
\item how many items per carousel ensure a good decision.
\end{enumerate*}

\heading{What types of carousel topics are the most helpful}
In our survey, we defined ten different carousel topics (based on \cite{loepp_multi-list_2023,10.1145/3301275.3302306} and examples taken from streaming services) for which we asked participants how useful on a Likert scale they have been or imagine they would be in finding a movie. %

Close to half of the participants (also the highest response among the 10) found the common \textit{genre} topics very helpful. \textit{Personalized} and \textit{item-based} were the next topics that also had many positive responses. \textit{Content}, \textit{user-based}, \textit{expert-based}, and \textit{temporal} topics were all found to be more helpful than not. Finally, \textit{global top/popular} was found to be slightly more helpful than not and the remaining two (\textit{regional top/popular}, \textit{exclusive}) were either neutral or not helpful. When comparing \textit{global top/popular} to \textit{regional top/popular}, it seems that participants would rather trust the global appeal of a movie than their own smaller region, but this may be mediated by culture and may vary by location. On the other hand, users preferred ``the smaller group'' personalized recommendations over user-based. 
In terms of unhelpful topics, most participants found that the \textit{exclusive} topic was not helpful in finding movies. %
While streaming service providers are incentivized to present their own shows and movies, it appears that it is not helpful for the user browsing experience to have an exclusive topic.

\begin{figure*}[!t]
    \centering
    \includegraphics[width=0.70\textwidth]{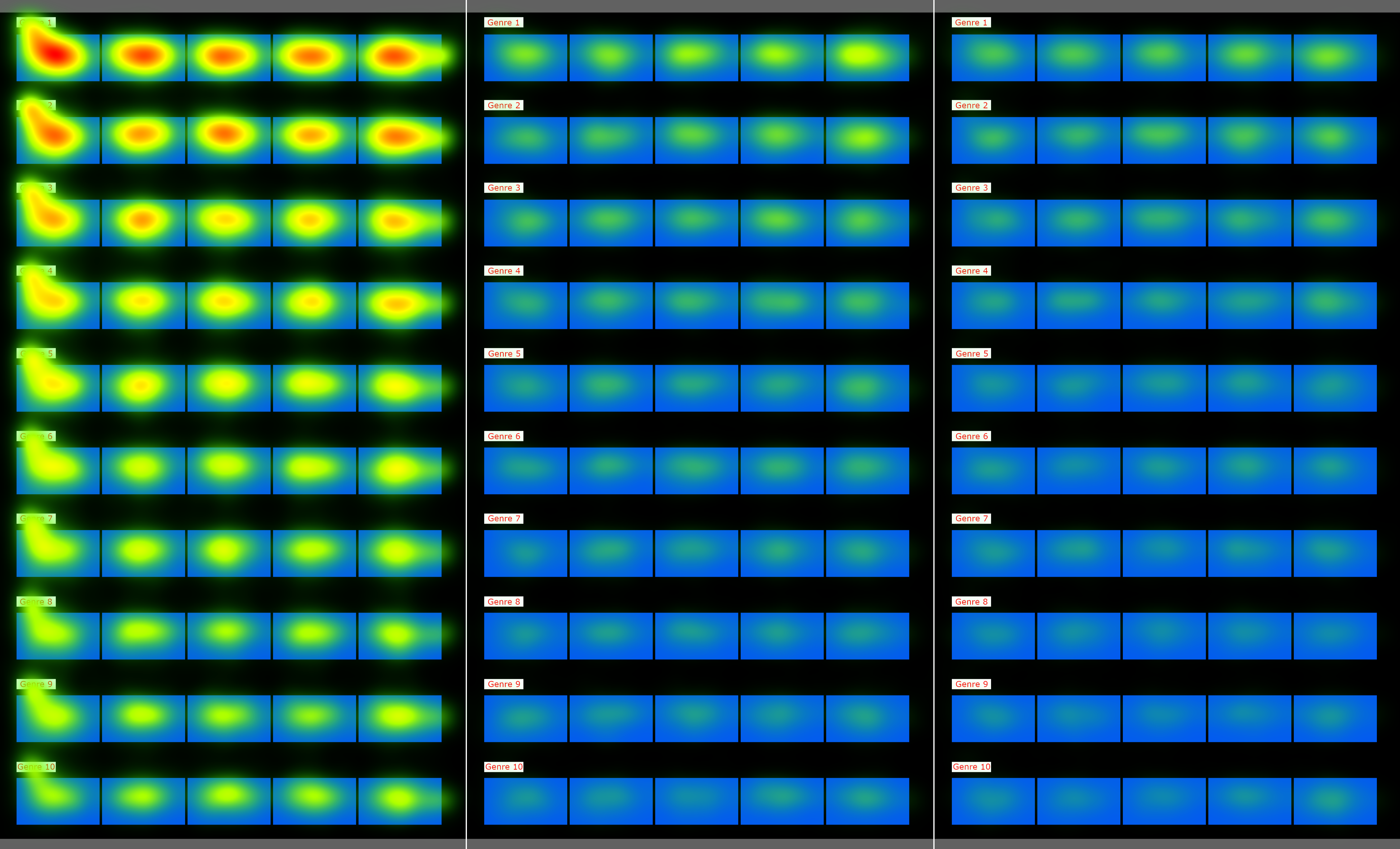} 
    \caption{Aggregate fixation heat map for every user on all 30 free-browsing screens with duration weighting shown on a to-scale background stimuli of the movie screens (with movie posters shown as blue and genre text as white boxes). It includes horizontal displacement, so initial 5 movies can be distinguished from second and third set from swiping.}
    \label{fig:fixation_subim_aggregate}
\end{figure*}
\label{survey most helpful topics}

\heading{How many carousels on a page}
\label{survey how many items}
In the post-task survey, participants were asked multiple questions for feedback on the design of the carousel, namely number of carousels and items. For the number of carousels, 27.6\% ($n=24$) of the participants answered that they felt overwhelmed by the 10 genre carousels presented. When asked about having more genre carousels on the page (additional to the 10 present), 2.3\% ($n=2$) would have liked to see 5 or more genres, 5.7\% ($n=5$) would have liked to see 3--4 more genres, 14.9\% ($n=13$) would have liked to see 1--2 more genres, and 77.0\% ($n=67$) found that number was sufficient. These results show that 10 (or slightly less) genre carousels could be a good amount to satisfy exploration needs, but not overwhelm. %

\heading{How many items in a carousel}
For the number of movies in the carousels, 19.5\% ($n=17$) of the participants answered that they felt overwhelmed by the 15 movies presented (5 with 2 swipes). When asked about having more movies, 8.0\% ($n=7$) would have liked to see 5--10 more, 9.2\% ($n=8$)  would have liked to see 15--20 more, 33.3\% ($n=29$) would have liked to see 25 or more additional movies, and 49.4\% ($n=43$) found that number was sufficient. In summary, 15 items (or more) seem to be a desirable number per carousel.
Interestingly, services like Prime Video or Netflix tend to have many more carousel genres displayed (with Prime Video having infinite scroll) and many more movies/shows in each carousel. Based on these results, users may seem to expect or be satisfied with the number of movies in each carousel on these services, but also may be overwhelmed by the large number of carousels.

\heading{Suggestions for improving the study interface}
\label{survey improving interface}
In addition to the above, participants were asked if they would have liked to see other types of carousels presented, of which 55.2\% ($n=48$) wanted more types in addition to the 10 genres, 27.6\% ($n=24$) wanted more types replacing some of the genres, and 17.2\% ($n=15$) said that the 10 genres were sufficient. %

Also, participants were asked to provide any comments or suggestions for improving the carousels with an open response. The most mentioned additions were: implementing IMDB ratings, adding director information, short clips/trailer on hovering on the poster, rearranging carousel topics to have preferred always at the top of the screen, and addition of different carousel topic types. The suggested features are found in popular streaming services and seem to be desired by users when interacting with a similar service.

\heading{Interest in customizable carousel homepages}
\label{survey customization}
One possible approach to improving the design of carousel interfaces is allowing users more autonomy by being able to customize their interface, which is in line with the suggestion to have preferred topics always at the top of the screen. %
Therefore, we asked the participants whether they would customize their streaming service homepage if they had the option to do so. Results showed that 70.1\% ($n=61$) would customize their homepage, 16.1\% ($n=14$) would not, and 13.8\% ($n=12$) did not know. This shows that allowing users to customize their homepage could be a feasible method to improve the user experience in carousel interfaces.

\subsection{Visualizing Browsing of Carousels}
\label{Visualizations}
\begin{figure*}[!t]
\centering
    \begin{subfigure}{0.49\textwidth}
    
     \includegraphics[width=0.99\linewidth]{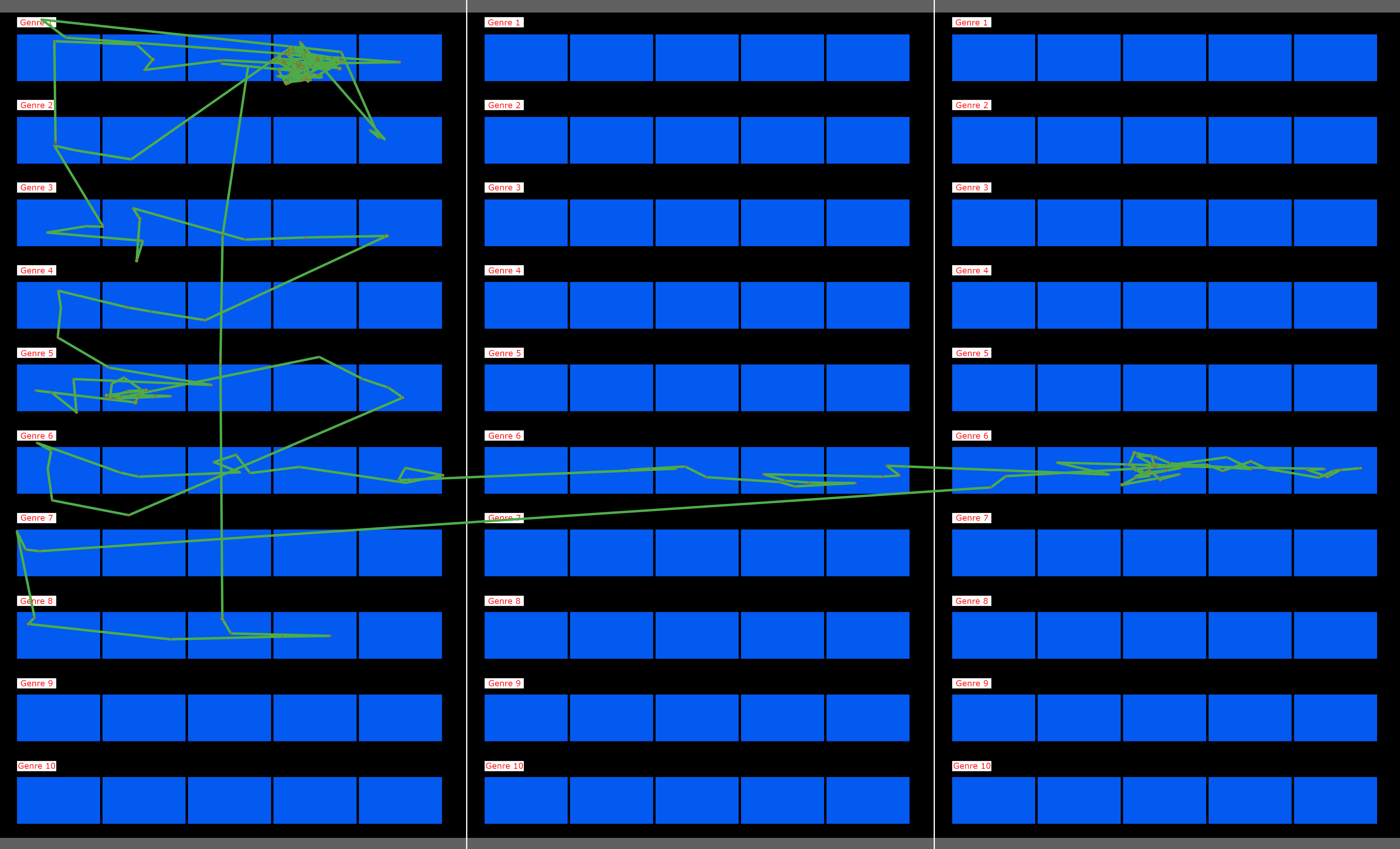}
     \caption{Fixation scan path with swipe differences}
     \label{fig:fixation_subim_fixation_scan}
    \end{subfigure}
    \begin{subfigure}{0.244\textwidth}
    
    \centering
    \includegraphics[height = 5.3cm]{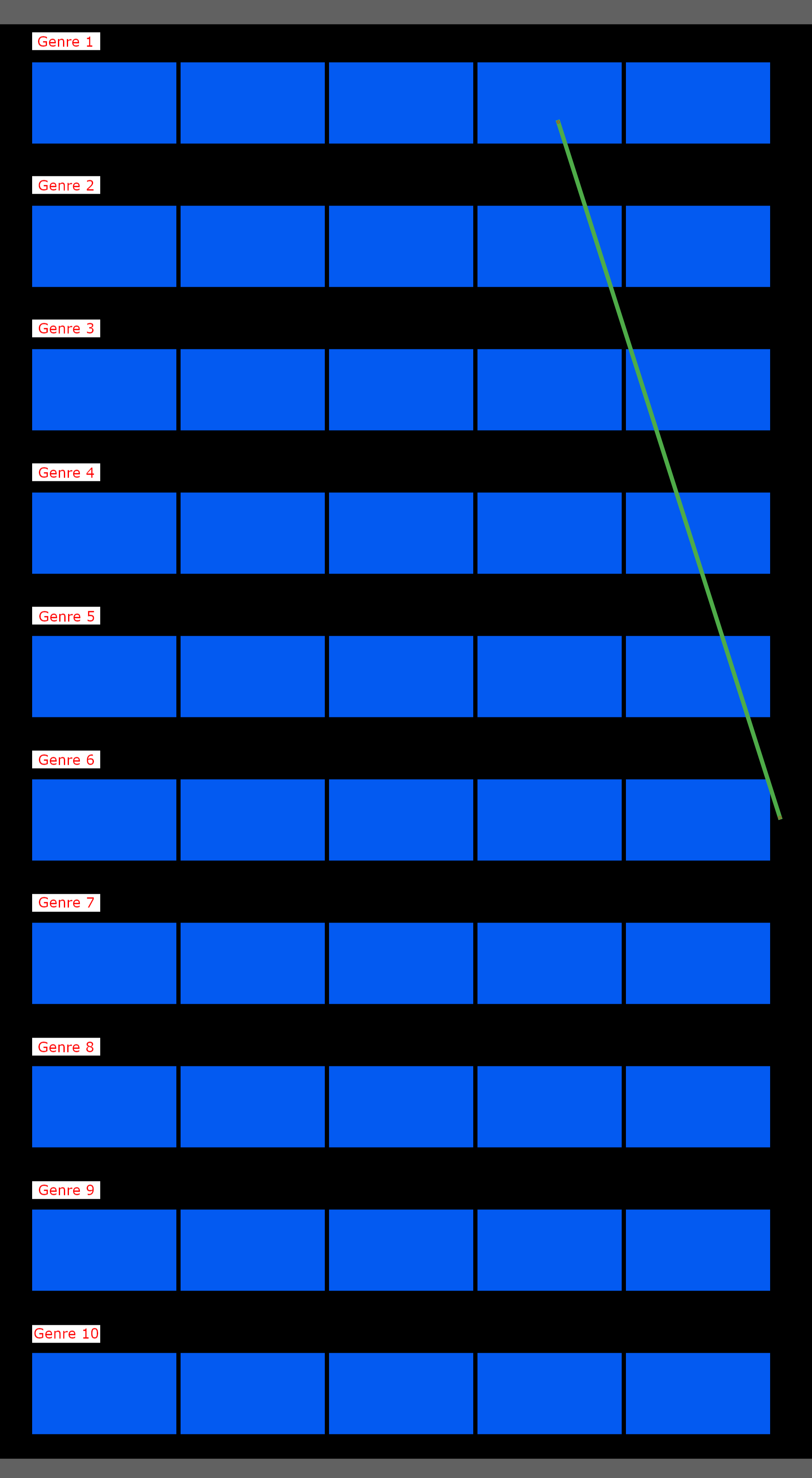}
    \caption{Click scan path}
    \label{fig:fixation_subim_click_scan}
    \end{subfigure}
    \begin{subfigure}{0.245\textwidth}

    \centering
    \includegraphics[height = 5.3cm]{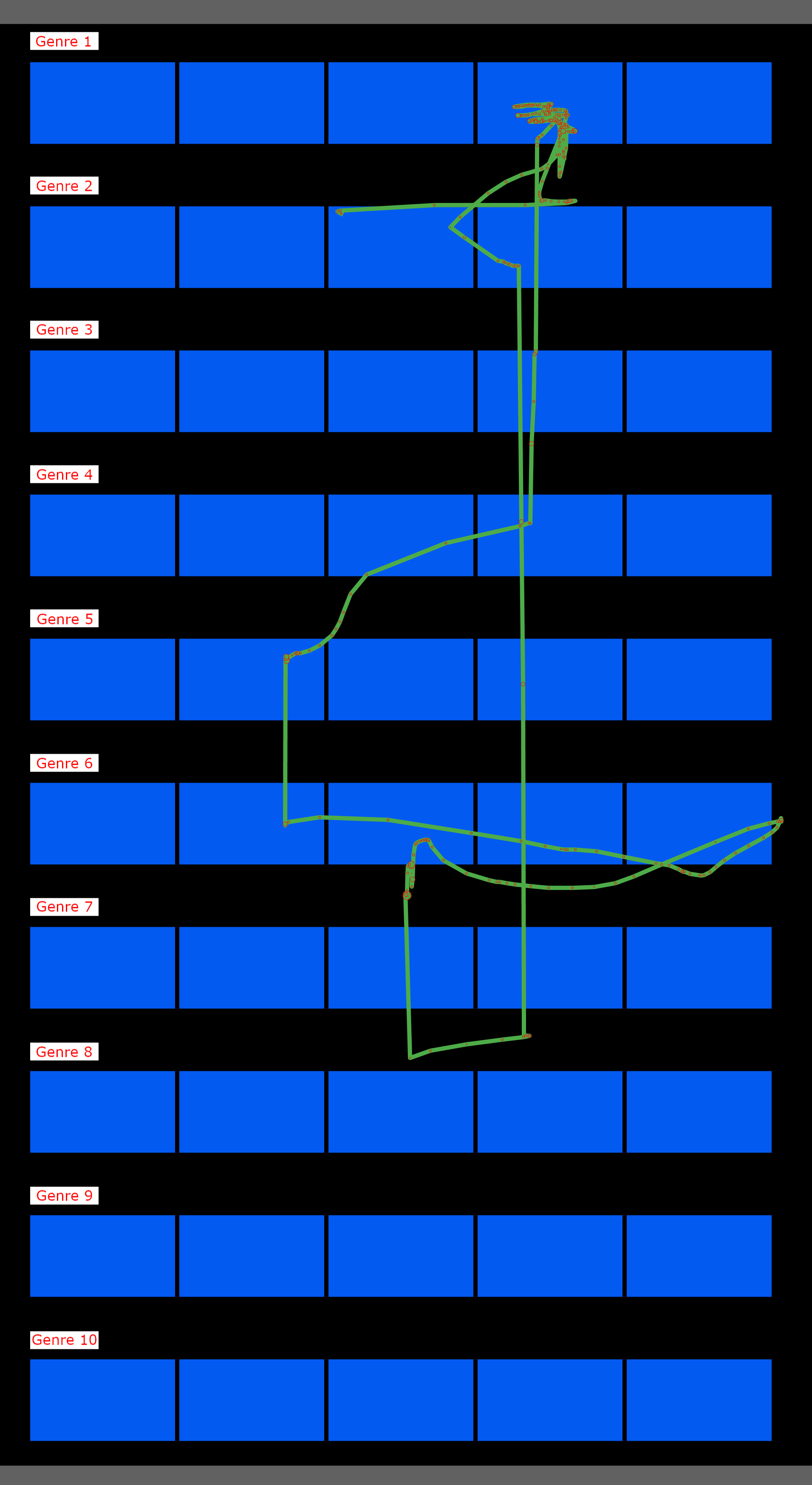} 
    \caption{Cursor scan path}
    \label{fig:fixation_subim_cursor_scan}
    \end{subfigure}
    \caption{Visualizations of fixations, clicks and mouse cursor movements for one selected user and one free-browsing screen.} %
    \label{fig:vis_fixation}    
\end{figure*}

In Figures~\ref{fig:fixation_subim_aggregate} and~\ref{fig:vis_fixation}, we present the first visualizations of gaze data in carousel interfaces. For \cref{fig:fixation_subim_aggregate,fig:fixation_subim_fixation_scan}, fixation data with $(x,y)$ pixel position value were used (non-public dataset), and \cref{fig:fixation_subim_click_scan,fig:fixation_subim_cursor_scan} used position value for clicks and cursor respectively (non-public dataset) rather than AOI allocation (public dataset).

In \cref{fig:fixation_subim_aggregate}, fixation data from all tasks across every user was combined to generate an aggregated heat map  on a to scale stimuli (also included in GitHub). The heat map also takes into account fixation duration. Therefore, the heat is representative of both how long and how often users were fixating on a location. The heat map points to a strong top-down browsing behavior, left-right browsing behavior (at least on page 1), and a  bias towards initially presented movies (unswiped). These preliminary results support the golden triangle browsing behavior or F-shaped pattern seen in single 2D  ranked lists \cite{10.1145/2959100.2959150}, which supports our original hypothesis that this behavior would also be seen in carousels \cite{10.1145/3616855.3635734}. However, on pages 2 and 3 more fixations can be seen on the farthest right movies which go against the general trend of F-pattern seen, particularly left-right browsing. We hypothesize that the increase in fixations on the last presented movie is due to how the swiping of the carousels works. However, there may be several different behaviors that could explain this result (e.g., central bias or a bias to the cursor/click position), which we plan to examine in our future work.  %

For \cref{fig:fixation_subim_fixation_scan,fig:fixation_subim_click_scan,fig:fixation_subim_cursor_scan}, we selected one participant and one free-browsing screen to visualize the types of data available. %
They present scan paths showing the points where data were registered (red circles) with radius proportional to the duration. For this screen, the participant began to fixate on the first, left-most movie of genre 2 and ended on the fourth movie of genre 1, as can be seen in the fixation data from \cref{fig:fixation_subim_fixation_scan}. %
\cref{fig:fixation_subim_click_scan} shows the clicks for two same location forward swipes on genre 6 and the final click to select the fourth movie of genre 1. Finally, the cursor path can be seen in \cref{fig:fixation_subim_cursor_scan} from an initial cursor position hovering the third movie of genre 2. The scan paths provide a detailed summary of the participant's browsing/interaction sequence. For example, multiple fixations (\cref{fig:fixation_subim_fixation_scan}) and mouse cursor movements (\cref{fig:fixation_subim_cursor_scan}) on the selected fourth movie of genre 1 are indicative of the user reading its description (tracing the text with the cursor). %

It is also possible to add user preference information to interpret these results. %
In \cref{fig:fixation_subim_fixation_scan}, we can see that the participant often examined multiple movies of each genre that they scrolled to. Outside of genre 1, the majority of fixations and all swipes are found in genres 5 (Drama) or 6 (Animation), which are both user's preferred ones. On the other hand, genre 4 (Fantasy), which was marked as the top genre by the user, lacked fixations when compared to the other ones. We provide this as an example and leave an analysis of the impact of genres on browsing as future work.

\section{Limitations}
\label{Limitations}
\heading{User study limitations} %
The hardware and software were identical between the two locations of the study, however, the lighting conditions differ, which %
could impact the accuracy of the gaze data.%

Another limitation is related to participants who may be skewed to those of above average socioeconomic level and education, which may not represent the average streaming service users. While this most likely impacts preferences for certain movies, it is unclear if it impacts how one would browse in general. %

\heading{Interface limitations}
The study did not explicitly consider poster visual bias. %
Color and content may %
impact browsing behavior. We address it only indirectly through aggregation (over screens/partici\-pants), in which the individual screen biases are less impactful.

Secondly, although we only gathered movie posters from the MovieDB API annotated as having English text, there were a few cases that did not have any text, which could possibly encourage (standing out from the rest) or discourage user interest. %

\heading{Dataset limitations}
Due to small inaccuracies in eye tracking, it is possible that some fixations are misattributed to incorrect locations/AOIs. This is most likely to happen when
\begin{enumerate*}[label=(\roman*)]
\item a participant is gazing at the border between 2 movies or 
\item moving from the genre text to the movie below. 
\end{enumerate*}
The second case could lead to attributing genre text AOIs to the movie below and vice versa.
An important limitation is the size of the dataset. While it is more than 9 times larger than the only available click feedback dataset in multi-list interfaces \cite{10.1145/3450613.3456809}, the 3,477 interactions provided in the dataset are still small when compared to industrial datasets. This may pose difficulties when training recommenders just on this data, but it is possible to connect the movies in our study with other datasets \cite{movielens}. 
Finally, results of the dataset may be limited to the movie domain. Browsing behaviors may vary in other domains, especially due to how different carousel topics can be to the movie domain.%

\section{Conclusion}
Although carousel and multi-list interfaces are widely used in commercial services, thus impacting users daily, there is little research that is focused on this type of interface. %
In this work, we help address the barriers to research in carousels and encourage a return to eye tracking and empirical understanding of user behavior by providing the RecGaze dataset. %
We demonstrated the utility of the dataset for a range of use cases and  provided the first gaze analysis of user behavior in carousel interfaces supporting the golden triangle or F-pattern browsing behavior. By providing this dataset, we hope to enable and advance research in carousel and multi-list interfaces as well as gaze-based recommendation systems. 

In future work, we will present an extensive analysis of the gaze data gathered herein for determining how users browse carousel interfaces taking into account preferences, interaction sequences (i.e., swipes), and decision-making scale results. 

\begin{acks}
We would like to acknowledge the support of Eye Square for providing hardware and software. Also, we would like to acknowledge Aneta Zugecova in helping administer the user study and Blickshift Analytics \cite{Blickshift_2024} for its support with software and eye tracking analysis. 
This work was supported by Eyes4ICU, a project funded by the European Union under the Horizon Europe Marie Sk\l{}odowska-Curie Actions, GA No.\ \href{https://doi.org/10.3030/101072410}{101072410}.
It is also partially supported by NWO under project numbers VI.Veni.222.269, 024.004.022, NWA.1389.20.\-183, and KICH3.LTP.20.006, and the European Union's Horizon Europe program under grant agreement No.\ 101070212.
All content represents the opinion of the authors, which is not necessarily shared or endorsed by their respective employers and/or sponsors.
\end{acks}

\balance

\bibliographystyle{ACM-Reference-Format}
\bibliography{references}
\end{document}